\documentclass[twocolumn,showpacs,preprintnumbers,amsmath,amssymb]{revtex4}

\usepackage{graphicx}
\usepackage{dcolumn}
\usepackage{bm}

\begin{document}

\preprint{APS/123-QED}

\title{Proof of the Standard Quantum Limit for Monitoring Free-Mass Position}

\author{Seiji Kosugi}
 \email{kosugi@jc.shukutoku.ac.jp}
\affiliation{%
Department of Food and Nutrition, Shukutoku Junior College, 6-36-4 Maenocho, Itabashi-ku, Tokyo 174-8631, Japan
}%

\date{\today}

\begin{abstract}
The measurement result of the moved distance for a free mass $m$ during the time $\tau$ between two position measurements cannot be predicted with uncertainty smaller than $\sqrt{\hbar \tau /2m }$.
This is formulated as a standard quantum limit (SQL) and it has been proven to always hold for the following position measurement: a probe is set in a prescribed position before the measurement. Just after the interaction of the mass with the probe, the probe position is measured, and using this value, the measurement results of the pre-measurement and post-measurement positions are estimated.
\end{abstract}

\maketitle

There has been considerable interest in the problem of whether fundamental quantum limits exist for repeated measurements of free-mass position, in particular for gravitational-wave detection. Braginsky and Vorontsov \cite{BV} derived a limit in the uncertainty $\Delta x$ in a second measurement result of the position of a free mass $m$ in a time $\tau$ after the first position measurement:
\begin{eqnarray}
   (\Delta x)_{\rm SQL} = \sqrt{\frac{\hbar \tau}{2m} }.
\end{eqnarray}
This is called the standard quantum limit (SQL) for monitoring the position of a free mass.

A recent controversy \cite{Yuen}-\cite{MOPRL} began with Yuen's argument \cite{Yuen} that the derivation given by Braginsky and Vorontsov contains a flaw.
In the Heisenberg picture, the position operator $\hat{x}_\tau$ of the free mass at time $\tau$ is given by 
$\hat{x}_\tau=\hat{x}_0+\hat{p}_0\tau /m $, 
where $\hat{x}_0$ and $\hat{p}_0$ are the position and momentum operators, respectively, at time $0$.
Thus, the variance of $x$ at time $\tau$ is given by
\begin{eqnarray}
   (\Delta x)^2 (\tau) = (\Delta x)^2 (0) + (\Delta p)^2 (0)(\frac{\tau}{m})^2  \nonumber  \\
   +(\langle \hat{x}_0\hat{p}_0 + \hat{p}_0\hat{x}_0 \rangle -2\langle \hat{x}_0\rangle \langle \hat{p}_0 \rangle)\frac{\tau}{m},
\end{eqnarray}
where $\Delta x (0)$ and $\Delta p (0)$ are the position and momentum uncertainties, respectively, at time 0.
The previous derivation of the SQL implicitly assumes that the correlation term [the last term in Eq. (2)] is zero or positive.
Yuen \cite{Yuen} has pointed out that some measurements leave the free mass in a contractive state, for which the correlation term is negative, and hence, the variance of $x$ decreases with time for a while. As a result, the variance $\Delta x(\tau)$ can be smaller than the SQL.
Note that Eq. (2) is about the intrinsic spreading of the free-mass wave function. In a measurement process, measurement errors are also crucial. This was emphasized by Caves, \cite{Caves} who showed that despite of Yuen's argument, the SQL holds true for a specific position-measurement model using von Neumann's interaction \cite{VNMF} 
\begin{eqnarray}
H=K\hat{x}_0\hat{P}_0
\end{eqnarray}
between the mass and probe, where $\hat{x}_0$ and $\hat{P}_0$ are the mass position and probe momentum, respectively, just before the interaction, and $K$ is a coupling constant.
The variance $\Delta_2$ of the result for the second position measurement is given by
\begin{eqnarray}
   \Delta_2 ^2 =\sigma ^2+ (\Delta x)^2 (\tau) \geq (\Delta x)^2 (0)+(\Delta x)^2 (\tau)      \nonumber  \\
   \geq 2\Delta x(0) \Delta x(\tau) \geq | \langle [\hat{x}_0,\hat{x}_\tau] \rangle| \geq \hbar \tau/2m,
\end{eqnarray}
provided that $\sigma \geq \Delta x(0) $, where $\sigma$ is an imperfect resolution of the measurement.
However, Caves did not prove that the inequality $\sigma \geq \Delta x(0) $ holds true in general, although it holds for von Neumann's interaction.
As pointed out by Ozawa, \cite{MOPRL} Caves's definition of the resolution is ambiguous.
He did not distinguish two kinds of measurement errors: an error $\epsilon$ in a measurement result for the mass position just before the measurement and an error $\sigma$ in a measurement result for the position just after the measurement.
These two errors are essentially different, although they are the same for von Neumann's interaction [Eq. (3)].
The errors $\epsilon$ and $\sigma$ were named the precision and the resolution, respectively, by Ozawa. \cite{MOPRL}

Caves formulated the SQL as follows: \cite{Caves} Let a free mass $m$ undergo unitary evolution during the time $\tau$ between two measurements of its position $x$, made with identical measuring apparatuses; the result of the second measurement cannot be predicted with uncertainty smaller than $\sqrt{\hbar \tau /2m }$ in average over all the first readout values.
Suppose that the mass and probe interact in the time interval $(0,t)$ at the first position measurement, and that the second measurement starts at $t=T$.
Then, Caves's formulation states
\begin{eqnarray}
   \int \Delta_{2,X} ^2 P(X) dX \geq \hbar \tau/2m, \nonumber \\
   \Delta_{2,X} ^2 =\epsilon(x_T)^2+ (\Delta x)_X ^2 (T),  
\end{eqnarray}
where $\tau=T-t$, and $P(X)$ is the probability distribution for obtaining the first readout value $X$ of the probe position.
The physical quantities $\Delta_{2,X}$ and $(\Delta x)_X (T)$ are distinguished by a subscript $X$ indicating that they are obtained when the readout value is $X$.
Caves did not take into account the $X$-dependence of the error $\epsilon(x_T)$.
$\Delta_{2,X}$ is the variance of the results of the second position measurement, and $\epsilon(x_T)$ is the measurement error for the mass position just before the second measurement.
Note that Caves's formulation does not include the measurement error $\sigma(x_t)$ for the first measurement result.

Ozawa \cite{MOPRL} demonstrated that a certain position measurement breaks the SQL formulated by Caves.
Using the interaction 
\begin{equation}
   \hat{H}=\frac{K\pi}{3\sqrt{3}}(2\hat{x}_0\hat{P}_0-2\hat{p}_0\hat{X}_0+\hat{x}_0\hat{p}_0-\hat{X}_0\hat{P}_0),
\end{equation}
where $\hat{p}_0$ and $\hat{X}_0$ are the mass momentum and probe position, respectively, just before the first measurement, and neglecting free Hamiltonians of the mass and probe, he obtained $\epsilon(x_T)=0$ for any initial state $|\phi_0 \rangle $ of the mass.
Moreover, when the initial state $| \xi_0 \rangle$ of the probe is taken to be a contractive state $|\mu, \nu, 0, \omega \rangle $, \cite{Yuen} the measurement leaves the mass in the contractive state $|\mu, \nu, X, \omega \rangle $ after the first position measurement. Thus, this measurement beats the SQL in the way suggested by Yuen.
On the problem of whether a SQL exists for repeated measurements of the free-mass position, Maddox \cite{Maddox} decided in favor of Ozawa.

However, several questionable points exist in his argument.
First, interaction (6) used by Ozawa does not conserve the total momentum of the mass and probe before and after the measurement, for the post-measurement momenta of the mass and probe are $\hat{p}_t=-\hat{P}_0$ and $\hat{P}_t=\hat{p}_0+\hat{P}_0$, respectively. \cite{MOPL1,MOPL2}
Because the disturbance in the mass caused by the first position measurement affects the position fluctuation $\Delta x(T)$ of the mass, the validity of his argument using such a interaction is questionable.
(Interaction (3), used by Caves, also does not conserve the total momentum. \cite{MOPL1,MOPL2})

Moreover, Ozawa \cite{MOPRL} supposed in his argument that the mean value of the readout of the probe position $\hat{X}_t$ just after the measurement is identical to the mean position of the free mass just before the measurement, and that the mean position of the free mass just after the measurement is identical to the readout value $X$, i.e.,
\begin{eqnarray}
   \langle \phi_0,\xi_0 | \hat{X}_t | \phi_0,\xi_0 \rangle &=&  \langle \phi_0| \hat{x}_0 | \phi_0 \rangle ,    \\
   X &=& \langle \phi_X | \hat{x}_0 | \phi_X \rangle ,
\end{eqnarray}
for all possible $| \phi_0 \rangle $, where $| \phi_X \rangle$ is the free-mass state just after the first measurement when the readout value is $X$, and we represent the tensor product $|\phi_0\rangle \otimes |\xi_0 \rangle$ as $|\phi_0, \xi_0 \rangle$.
Because the right-hand side of Eq. (8) equals $X- \langle \hat{X}_0 \rangle $ for interaction (6), assumption (8) can be justified only when the average $ \langle \hat{X}_0 \rangle $ of the initial probe position is $0$.
In fact, Ozawa adopted the initial state $| \xi_0 \rangle $ of the probe that satisfies $ \langle \hat{X}_0 \rangle =0$.
However, this means that assumption (8) is satisfied only in the specified coordinate system in which an observer has $ \langle \hat{X}_0 \rangle =0$.
In general, assumption (8) cannot be justified.

As a position measurement that breaks SQL (5) formulated by Caves, we propose the following interaction:
\begin{equation}
   \hat{H}=K(\hat{p}_0+\hat{P}_0)(\hat{X}_0-\hat{x}_0),
\end{equation}
which conserves the total momentum of the mass and probe independently of the value of $g_0 \equiv Kt$.
For this interaction, neglecting the free Hamiltonians of the mass and probe, we obtain
\begin{equation}
   \hat{x}_t=(1-g_0)\hat{x}_0+g_0\hat{X}_0, \quad \hat{X}_t=-g_0\hat{x}_0+(1+g_0)\hat{X}_0.
\end{equation}
When interaction (9) with $g_0=-1$ is used, we obtain $\epsilon(x_0)=0$ for any $| \phi_0 \rangle$.
Moreover, when the initial state of the probe is taken to be the contractive state $|\mu, \nu, \langle \hat{X}_0 \rangle, \omega \rangle $, the measurement leaves the mass in the contractive state $|\mu, \nu, 2X-\langle \hat{X}_0 \rangle, \omega \rangle $ after the first position measurement.
Therefore, this measurement beats SQL (5) formulated by Caves.
Note that this measurement does not include the flaws in that proposed by Ozawa and that it does not break SQL (11) mentioned below.

The arguments hitherto made are based on the SQL formulated by Caves.
Here, we reexamine the formulation of the SQL for repeated measurements of free-mass position.
Braginsky and Vorontsov \cite{BV} introduced the SQL when deriving the SQL for measuring the constant force $F$ acting on a free mass.
Using a change of position $\delta x=F\tau^2/2m$ caused by the force $F$, they proved that the SQL for measuring the force $F$ is equal to $2m (\Delta x)_{\rm SQL}/\tau ^2$, where $\tau$ is the duration of action, and $(\Delta x)_{\rm SQL}$ is the SQL for the free-mass position.
Caves et al. also developed the same argument on page 359 in Ref.(13).
Moreover, Braginsky and Khalili \cite{BK} derived the standard quantum limit $(\Delta F)_{\rm SQL}$ for the mass sensitivity to the force using the standard quantum limit $(\Delta P)_{\rm SQL}$ for momentum.
In their arguments, $(\Delta P)_{\rm SQL}$ was considered to be the uncertainty of the result $m(x_2-x_2)/\tau$ obtained using the results $x_1$ and $x_2$ of the two position measurements.
For these derivations to be justified, $(\Delta x)_{\rm SQL}$ must be the uncertainty in the moved distance obtained from the results of two successive position measurements.

Based on the above discussion, we formulate the SQL as follows: Let a free mass $m$ undergo unitary evolution during the time $\tau$ between two measurements of its position $x$, the measurement result of the \textit{moved distance} cannot be predicted with uncertainty smaller than $\sqrt{\hbar \tau /2m }$.
In our formulation, it is not necessary that the two position measurements must be made with identical measuring apparatuses.
Averaging the uncertainty over all the first readout values is also unnecessary.
Although in Caves's formulation the measurement error $\sigma(x_t)$ for the first position measurement is not included, it must be taken into account in our formulation:
\begin{eqnarray}
   \Delta_X ^2 = \sigma_X (x_t)^2+\Delta_{2,X} ^2  \geq \hbar \tau/2m, 
\end{eqnarray}
where $\Delta_{2,X}$ is the uncertainty of the result for the second position measurement. It is given for interactions (3) and (6) by 
\begin{eqnarray}
\Delta_{2,X} ^2= \epsilon_{X^{\prime}} (x_T)^2+(\Delta x)^2 _X (T), 
\end{eqnarray}
where $X^{\prime}$ is the readout value of the second position measurement and $(\Delta x)_X (T)$ is the variance of the mass position just before the second measurement.
It is not clear that Eq. (12) always holds, although it holds for interactions (3) and (6).
However, a measurement theory should correctly reproduce the Born rule of probability for any object state $| \phi_0 \rangle$, when the measurement error $\epsilon(x_T)$ is zero.
Then, $\Delta_{2,X} = (\Delta x)_X (T)$. Therefore, it is reasonable to assume that the following inequality always holds:
\begin{eqnarray}
   \Delta_{2,X} \geq (\Delta x)_X (T).
\end{eqnarray}
Thus, if
\begin{eqnarray}
\sigma_X (x_t) \geq (\Delta x)_X(t),
\end{eqnarray}
then 
\begin{eqnarray}
   \Delta_X ^2 &\geq& \sigma_X (x_t)^2+ (\Delta x)^2 _X (T) \geq (\Delta x)^2 _X (t)+(\Delta x)^2 _X (T) \nonumber  \\
   &\geq& |\langle [\hat{x}_t,\hat{x}_T] \rangle | \geq \hbar \tau/2m.  \nonumber
\end{eqnarray}
Therefore, when inequality (14) holds, the SQL is valid even though the free mass after the first measurement is in a contractive state.
Contractive states do not vitiate this argument, as pointed out by Caves.

We now prove that inequality (14) always holds for the position measurement according to the following method: the probe is arranged in a prescribed position $ \langle \hat{X}_0 \rangle $ with a small position fluctuation before the measurement. Just after the interaction of the mass with the probe, the probe position $\hat{X}_t$ is measured using another apparatus, and using this value, the measurement results of the mass positions $\hat{x}_0$ and $\hat{x}_t$ are estimated.
It is assumed that the probe position $\hat{X}_t$ can be precisely measured.
The detailed discussion about our position-measurement model is found in Ref. (15).
Let $\hat{U}$ be a unitary operator representing the time evolution for the first position measurement.
Then, the state of the system composed of the mass and probe after the interaction should be given by
\begin{eqnarray}
   \hat{U}| x_0, X_0 \rangle = | f(x_0, X_0), g(x_0, X_0) \rangle ,
\end{eqnarray}
where $| x_0, X_0 \rangle $ is an eigenstate of the position operators $\hat{x}_0$ and $\hat{X}_0$, and $f(x_0, X_0)$ and $g(x_0, X_0)$ are arbitrary functions of $x_0$ and $X_0$.

From the law of conservation of the linear momentum, we obtain
\begin{eqnarray}
  x_t \equiv f(x_0, X_0) = x_0 + F(x_0-X_0),  \\
  X_t \equiv g(x_0, X_0) = X_0 + G(x_0-X_0).
\end{eqnarray}
where $G(x)$ and $F(x)$ are arbitrary functions.
Because the value of $x_0$ must be determined uniquely using the value of $X_t$ for arbitrary possible value of $X_0$, the real number $X_t$ must have a one-to-one correspondence to $x_0$. Then, there exists an inverse function $G^{-1}(x)$. Thus, we obtain
\begin{eqnarray}
  x_0 = X_0 + G^{-1} (X_t-X_0).    \nonumber
\end{eqnarray}

The spreading of the probability density for the probe causes the errors of position-measurement results $(x_0)_{\rm exp}$ and $(x_t)_{\rm exp}$.
We cannot know which position component $X_0$ of the probe wave function interacts with the mass.
Therefore, we define the measurement results $(x_0)_{\rm exp}$ and $(x_t)_{\rm exp}$ as the values of $x_0$ and $x_t$, respectively, obtained when $X_0$ is set equal to $\langle \hat{X}_0 \rangle $:
\begin{eqnarray}
  (x_0)_{\rm exp} &=& \langle \hat{X}_0 \rangle + G^{-1} (X_t-\langle \hat{X}_0 \rangle),  \\
  (x_t)_{\rm exp} &=& (x_0)_{\rm exp} + F((x_0)_{\rm exp}-\langle \hat{X}_0 \rangle).
\end{eqnarray}
From the above equations, the measurement result $(x_t)_{\rm exp}$ for the post-measurement position is found to be a function of $X_t$ and $\langle \hat{X}_0 \rangle $.

Using the wave function $\langle x,X | \hat{U} | \phi_0,\xi_0 \rangle$ of the system just after the first measurement, we obtain the mass wave function obtained when the readout value is $X$:
\begin{eqnarray}
   \langle x | \phi_X \rangle &=& \langle x,X | \hat{U} | \phi_0,\xi_0 \rangle / \sqrt{P(X)},   \\
   P(X) &=& \int |\langle x,X|\hat{U}|\phi_0,\xi_0 \rangle |^2 dx.
\end{eqnarray}
Thus, the variance of position in the mass state $| \phi_X \rangle $ is
\begin{eqnarray}
(\Delta x)_X^2(t)  &=& \int (x-\langle \hat{x} \rangle _X)^2|\langle x | \phi_X \rangle |^2dx,   \\
\langle \hat{x} \rangle _X &=& \int x|\langle x | \phi_X \rangle |^2dx.
\end{eqnarray}

Since the probability density for obtaining the measurement results $x$ and $X$ for the post-measurement positions  $\hat{x}_t$ and $\hat{X}_t$, respectively, is $|\langle x,X|\hat{U}|\phi_0,\xi_0 \rangle |^2$, the square of the measurement error $\sigma(x_t)$ is given by 
\begin{eqnarray}
 \sigma(x_t)^2 = \int \{ (x_t)_{\rm exp} -x \}^2 |\langle x,X|\hat{U}|\phi_0,\xi_0 \rangle |^2 dx dX.
\end{eqnarray}

Then, we obtain the square of the measurement error $\sigma_X (x_t)$ for the post-measurement position $\hat{x}_t$ obtained when the readout value is $X$:
\begin{eqnarray}
\sigma_X(x_t)^2 = \int \{(x_t)_{\rm exp}-x \}^2|\langle x | \phi_X \rangle |^2dx.
\end{eqnarray}
Using the fact that $(x_t)_{\rm exp}$ is a function of $X$ and $\langle \hat{X}_0 \rangle $, and does not depend on $x$, we obtain
\begin{eqnarray}
\sigma_X (x_t)^2 &=& (\Delta x)^{2}_{X}(t)  + (\langle \hat{x} \rangle _X-(x_t)_{\rm exp})^2 \nonumber \\
&\geq& (\Delta x)_X^2(t) .
\end{eqnarray}
Therefore, SQL (11) always holds true.

\end{document}